# Class-Specific Data Augmentation: Bridging the Imbalance in Multiclass Breast Cancer Classification

Mahammadli Kanan[1], Bereketoğlu Abdullah Burkan[2], and Kabakci Ayse Gül[3]

[1] kanan.mahammadli@metu.edu.tr – Department of Mathematics, Middle East Technical University

[2] burkan.bereketoglu@metu.edu.tr – Department of Physics, Middle East Technical University

[3] akabakci@cu.edu.tr – Faculty of Medicine, Cukurova University

*Abstract*—Breast Cancer, the most common cancer among women, which is also visible in men, and accounts for more than 1 in 10 new diagnoses of cancer each year. It is also the second most common cause of women who die from cancer. Hence, it necessitates early detection and tailored treatment. Early detection can provide appropriate and patient-based therapeutic schedules. Moreover, early detection can also provide the type of cyst. This paper employs class-level data augmentation, addressing the undersampled classes' raising their detection rate. This approach suggests two key components: class-level data augmentation on structure-preserving stain normalization techniques to hematoxylin and eosin-stained images and transformer-based ViTNet architecture via transfer learning for multiclass classification of breast cancer images. This merger enables categorizing breast cancer images with advanced image processing and deep learning as either benign or as one of four distinct malignant subtypes by focusing on class-level augmentation and catering to unique characteristics of each class with increasing precision of classification on undersampled classes, which leads to lower mortality rates associated with breast cancer. The paper aims to ease the duties of the medical specialist by operating multiclass classification and categorizing the image into benign or one of four different malignant types of breast cancers.

*Index Terms*— Breast Cancer, Class Level Data Augmentation, Vision Transformers, Transfer Learning, Stain Normalization

## 1 INTRODUCTION

CANCER disease is the leading illness that causes death worldwide today. WHO (World Health Organization) estimates nearly 10 million deaths from cancer in 2020 in an article written in 2022 [1]. WHO also states that the most common cancers are breast, lung, colon-rectum, and prostate [1]. Furthermore, WHO adds that 85% of breast cancer arises from ductal cancers, and the rest is from lobules in the glandular tissue of the breast [2]. Male breast cancer is between one in a hundred and five in a thousand range; from a medical perspective, it is a rare disease for males but widely typical among female humans [2].

Breast cancer is the most common cancer diagnosed among women, accounting for 1 in every 10 women diagnosed yearly [3].

Furthermore, for women, breast cancer is the second most common death cause of cancer worldwide [3]. Breast cancer is a silent killer that slowly spreads and, for that reason, needs routine screening by the patient and a medical specialist on the case [3]. Diagnosis of breast cancer in women relies on specific tests. With improved state-of-the-art machine learning methods, specialists can have better quality and earlier detection time of the malignant cyst or cancer of the patient [3]. Although there exist several identification diagnostic techniques for breast cancer, and they are also improved with machine learning techniques to help the specialist diagnose breast cancer early and clearly, diagnosis by histopathology tissue images can be stated as the standard diagnosing technique for almost all types of cancers which also includes breast cancer [4]. In histopathology, tissue sample staining methods are used to detect the cancer tissue in the model [5]. The most fundamentally accepted staining technique on histopathology samples taken as the core image color is the H&E, which is hematoxylin and eosin staining that gives a pink color to the histopathology tissues at the end [5]. Pathologists can easily detect anomalies in the tissue with hematoxylin and eosin staining due to cancerous tissue; there, a non-uniform color spreads and makes cancer cells visible by the binding abilities of H&E [5]. In recent years, deep learning has outperformed other machine learning algorithms in computer vision tasks, as well as in medical imaging, such as binary and multiclass classification ([6], [7]), detection [8], and segmentation [9]. Introducing Convolutional Neural Network-based models in 2012, with Alexnet, the winner of the ImageNet classification, has played a key role in improving CNNs for computer vision. Inspired by Alexnet [10], other CNN architectures have also been developed over

---

Mahammadli K. *(Corresponding Author)* Author is with the Department of Mathematics, Middle East Technical University, Ankara, Turkey (e-mail: kanan.mahammadli@metu.edu.tr).

Bereketoğlu AB. Author is with the Department of Physics, Middle East Technical University, Ankara, Turkey (e-mail: burkan.bereketoglu@metu.edu.tr).

Kabakci AG. The author is with the Department of Anatomy, Faculty of Medicine, Cukurova University, Adana, Turkey (e-mail: akabakci@cu.edu.tr).

the years, such as VGG-16 [11], GoogLeNet (Inception) [12], ResNet [13], DenseNet [14], MobileNet [15]. Using these models via transfer learning, in other words, taking pre-trained weights of CNN architectures on Imagenet and fine-tuning on other custom data, is a common strategy in medical imaging to avoid overfitting and providing high-scored models. Current state-of-the-art breast cancer classification models on the BreakHis dataset [16], which is also the dataset used in this research, rely on pre-trained CNN models ([17], [18], [19], [20], [21]). In 2017, with the Transformers [22] introduction, a new stage began in the deep learning era. Then, Vision Transformers [23] was also shown, surpassing the CNN models on the classification tasks. By inspiring the rise of Vision Transformers, we suggest ViTNet architecture for multiclass classification of breast cancer on the BreakHis dataset by adapting class-level regularization and analysis.

*Our contribution can be summarized as follows*
1) Applying class-level data augmentation for the first time on the BreakHis dataset and overcoming the undersampling problem; 2) Constructing Transformer based automatic detection - ViTNeT for cancer classification; 3) Providing class-level scores to check the model's bias towards each class, preventing reliance on overall metrics bias; 4) Showing that on class level approaches, ViTNeT advances the results of current SOTA CNN based models by a max of 28 % in precision, 43 % in the recall, 24 % in f1-score and on overall metrics it surpasses the best CNN models by 21 % in accuracy, 15 % in macro averaged precision, 22 % in weighted averaged precision, 21 % in both macros averaged recall and weighted averaged recall, 19 % in macro averaged f1-score and 21 % in weighted averaged f1-score

This paper is organized in the following sequence. Section II will discuss the Related Works and continue our approach and methodology in Section III. Furthermore, we will provide results in Section IV and finish up with a conclusion in Section VI.

## 2 RELATED WORKS

### A. Mammography-Based Identification

Mammography is one of the main diagnostic techniques for mid-aged women to detect breast cancer or the lump type in their breasts ([24], [25]). It is known that regular mammographic imaging for women decreases the chance of rapid and aggressive growth of the tumors in the breast ([24], [26]). Cancerous tumors start at the breast's ducts and spread to or from the milk-producing glands or lobules; they must detect tumors early for higher chances of curing cancer ([24], [27]). Most medical doctors use mammography to diagnose whether the cyst is benign or malignant before taking a patient's biopsy and looking at the histopathological tests [25]. This diagnostic without CAD (computer-aided diagnostics) with machine-learning/Artificial-Intelligence/Deep-learning methods or with CAD only takes heaps of time that is needed to start the therapy earlier and has more chances of saving most tissue and life ([24], [26], [27]).

Medical experts in the past successfully identified, as mentioned in the previous paragraph, whether a breast cyst is benign or malignant, as their subclasses if the cyst is malignant with chemical coloring, physical shape, curve, and size tests ([24], [26], [27], [28]). These tests consume more time than recent studies on histopathological and mammography image classification systems with deep learning and machine learning techniques [28]. These new techniques with SVM and Naive Bayes are used to achieve accuracy up to 100% with some faulty occurring, such as Type-1 error of misclassifications occurring that is not possible with regular tests ([24], [28]). Mammography testing is highly successful in the benign and malignant classification accuracy performance metric. However, no subclass differentiation could be achieved with false-positive errors, even with newer models ([27], [28]). Furthermore, prior studies show that using breast density in mammographic breast cancer analysis with deep learning models, which are mammography-based deep learning models that look for mammographic density that increases accuracies of breast cancer risk models, shows us that with densities at specific locations of healthy and sick patient breasts on mammographic images, with the density-based analysis increases the accuracy, therefore, the chance of early detection of the tumor itself [25].

### B. WSI-Based Classification

Medical images for breast cancer classification can be in different formats, like DM, DBT, US, MRI, and Pathology. However, it has been the best practice to use histopathology images for computer-aided image analysis and deep learning techniques, which provide better insights into the cancer cells and help localize the exact places of metastasis. Therefore, medical pathology experts have developed techniques for achieving high-resolution histopathology images. The most advanced one is WSI - Whole Slide Images, made by digitalizing the glass sliding at a microscopic level with different magnifications. However, working with the Whole Slide Images is very expensive and requires heavy computational power and processing. On average, WSI size is around 40000x4000 pixels and can reach 100000x100000 [29].

One groundbreaking research in applying deep learning for Whole Sliding Breast Cancer Images is "Deep Learning for Identifying Metastatic Breast Cancer," proposed by Wang et al.[30], That won the Camelyon Grand Challenge 2016 - automatic detection and localization of metastatic breast cancer held by the International Symposium on Biomedical Imaging. 270 WSI images have been used for training and preprocessed before feeding into the network. Training images have been patched into 256x256 pixel-sized images with different magnifications: 40x, 20x, and 100x. Then, AlexNet [10], VGG16 [11], GoogLeNet [12], and FaceNet [30] architecture were used to train and test 230 WSI images.

During the training, patch-based classification was used. The algorithm identifies the breast cancer based on the given patch by looking for whether the offered patch is located in the tumor area. Used AUC score and Accuracy metrics to test the results, and GoogLeNet and VGG16 led to the best accuracy of 98.4 and 97.9%. Then, extracted features from Convolutional Layers were used as post-processing to produce heatmaps, and 28 geometrical and morphological characteristics were selected for slide-based classification and achieved an AUC of 0.925.

K. Das et al.[29] have used the Deep Multiple Instance CNN model to handle the computation power concerning high-resolution input images. Avoiding heavy computations, the authors have adopted Multiple Instance Learning with a Multi-Instance Pooling Layer - called MIPool, to aggregate the produced features by Convolution Layers. Bag-level annotation has been used instead of the individual, instance-level annotation for the training; however, the error is calculated by instance-level Margin Loss to keep the effect of single instances. Based on the results, the paper achieves a 94.44 % recall score on 40x and 200x magnified WSI images and 96.89 % and 90.48 on IUPHL and UCSB cancer datasets, respectively. However, the proposed architecture surpasses other Multiple Instance algorithms mainly on accuracy score.

Besides the publicly available datasets, some researchers collect the WSI breast cancer images by contacting the hospitals,s. Mi et al. [31] have acquired the Hemotoxyling and Eosin (H&E) stained raw WSI dataset from PUMCH. The total size of 540 images is approximately 706 GB. Since the dataset consists of raw images, manual patching of the image slices has been cropped for training the network with small image sizes and data augmentation. The authors use ImageNet-based pre-trained Inception-v3 by adding a fully connected layer of 1024 neurons with a softmax classifier as a patch-level classifier, reaching 86.67 % overall accuracy. Besides the fully connected layer, different Machine Learning models have also been used to classify breast cancer by feeding feature maps from Inception-v3. Among the algorithms - AdaBoost, Decision Tree, SVM, Random Forest, Gradient Boosting, LightGBM, and XGBoost, the last one reaches the highest accuracy of 90.43 %.

Even though Whole Sliding Images are great because of the high resolution and better view of the cancer cells and metastasized areas, WSI datasets come in substantial sizes. Therefore, keeping the dataset, preprocessing the training images, and training a deep neural network requires expensive resources. If the methodology of most researchers considered, as they mostly rely on the pre-trained models with millions of parameters and large size because of the saved weights, training them further on these images leads to further cost. As a result, the produced model also becomes a heavy and crucial factor of medical imaging research: keeping the trained models and using them on devices for automatic classification and detection of cancers. However, deploying these heavy models is impractical and requires further network optimization.

*C. Multiclass Classification from Histopathological Images*

As Han et al. [32] suggested in their work in 2017, multi-classification plays an essential role in reducing the workload on histopathologists. It enhances the chance to start a more specific treatment on the patients since it provides additional information about the situation. The work also uses the BreaKHis dataset [16], which is planned to be used in this work. They claim that one of the two main challenges of multi-classification on medical images is the variety in the sample images' resolutions and color distributions. To overcome this problem, instead of using general feature descriptors in feature extraction steps such as Gray-Level Co-occurrence Matrix (GLCM) and Parameter-Free Threshold Adjacency Statistics (PFTAS). They have decided to use a state-of-the-art CNN model, a Class Structure-based Deep Convolutional Neural Network (CSDCNN), which performs the feature extraction step on the dataset by end-to-end learning. Their experimental results show that this approach performs 13.4-14.7% better than traditional feature descriptors.

Due to clinical regularity, the dataset consists of eight imbalanced classes due to clinical frequency, where each of the four belongs to benign and malignant categories. As a solution, the work utilizes over-sampling by intensity variation, rotation, flip, and translation and tries to avoid data imbalance and over-fitting problems.

They also added some constraints by prior knowledge to discard the similarities between subclasses in the feature space. The paper indicates that not using a data augmentation step, developing the model on raw data is insufficient for CNN. The dataset consists of histopathological images from 82 patients, and they have tried separating each patient's images as training, validation, and test images. The overall results show no significant improvement in developing the model based on image or patient level. The work has tried training CSDCNN from scratch. However, using the weights from transfer learning gave ~3% better results. The work has achieved 93.2% accuracy with their additional methods on CSDCNN.

*D. Classification Disadvantages of Prior Methods*

Even though different medical imaging types like mammography and CT scans can be used for breast cancer classification, histopathology images provide better features for deep learning algorithms since they are a microscopic view of the cancer cells and their surrounding tissues and allow direct analysis of cancer. However, original high-resolution histopathology images - WSI is costly and requires vast analysis resources [31]. Therefore, we use low-resolution and small-sized image dataset BreaKHis not to deal with heavy computations. So far, researchers working on the BreaKHis dataset mainly rely on pre-trained CNN models and take advantage of the one or more fully connected layers with softmax by freezing a different number of layers from the pre-trained model. Furthermore, models resulting in high accuracy, trained for binary classification of malignant or benign, lead to worsened predictions when standard breast images are added to the training [33].

Even though Transfer Learning via CNN models can produce high scores, some factors still have not been considered yet for cancer detection. These great results can be misleading, and methods for handling that kind of situation have not been discussed so far. Most of the research done on the BreakHis dataset has been towards to binary classification ([17], [18], [34], [35]. 36], [37], 38], [39], [40], [41], [42], [43], [44], [45]). Some papers use accuracy as an evaluation metric, or the score that has been improved is the accuracy ( [28], [37], [42], [43], [46], [47], [48]). Although multiclass classification ([7], [19], [21]. [22], [49], [50], [51], [52], [53], [54], [55], [56], [57], [58]) has also been done on the BreakHis dataset, they do not provide scores across each class, but overall accuracy, precision or recall. In some papers, we also detected issues related to the statistics provided which seems unrealistic compared to the real-world scenario where there is always exist some uncertainty. For example, having 100 % specificity [57], or 100 % precision, recall, f1-score, and AUC score [54], over 99 % average specificity, accuracy, precision and recall ([53], [56]), data leakage [55], high overfitting - validation accuracy around 70 % while accuracy on training data over 90 % during the training [7]. Lastly, all of the papers ([7], [17], [18], [19], [21], [22], [28], [34], [35], [36], [37], [38], [39], [40], [41], [42], [43], [44], [45], [46], [47], [48], [49], [50], [51], [52], [53], [54], [55], [56], [57], 58]) use the same data augmentation for all the classes, and have not examined different techniques based on the breast cancer type, however it has been proved that when same augmentation methods applied to all the classes, majority classes which are called "label preserving" get positively effected, while minority class data become noisy and add bias to the model [59]. This situation makes underrepresented classes worsen, however as majority classes take advantage of augmented data, even from images coming from other classes, misleading improved overall scores appear at the end [59].

Considering all these issues, we provide automatic multiclass detection of breast cancer using ViTNet with class-specific data augmentation and summarize the overall results, as well as for each class separately.

Section III will delve more into the proposed approach and how it overcomes the issues discussed in Section II.

### 3 METHODOLOGY
*A. BreakHis Dataset*

The BreakHis dataset consists of breast cancer (BC) histopathology images that are collected from a clinical study of all patients in collaboration with the P & D Laboratory, Pathological Anatomy and Cytopathology, Parana, Brazil, and the Federal University of Paraná, informatics laboratory ([16], [60]). The samples in the dataset are generated from the breast tissue's biopsy surface, collected by surgical biopsy, and labeled by pathologists [60]. The preparatory process of the histopathology samples is the standard paraffin process regularly used as a common practice among clinicians, allowing researchers to view preparations via a microscope [60].

The BreaKHis dataset is divided into two main categories: benign and malignant tumors [16]. In histology terms, benign refers to a lesion that does not match any criterion of malignancy, e.g., marked cellular atypia, mitosis, metastasizing, etc. [16]. In standard cases, benign tumors are considered "innocents," which present slow-growing and localized lesions/cysts [16]. On the other hand, a malignant tumor can be referred to as the synonym for cancer, a lesion that can invade and destroy adjacent structures [16]. This characteristic of malignant corresponds to its being locally invasive, spreading to distant localizations, which is a synonym for metastasizing and causes death [16].

The pathologists implement the collection period of the images at different magnifications as follows [24]. First, the pathologist identified the region of interest (ROI) and the tumor [24]. The pathologist took some images with 40x magnification resolution to get complete coverage of the entire ROI [24]. Pathologists also prefer to select images with a tumor of a single type (in most cases, for better classification) [24]. Still, some photos may also consist of transitional tissue, for example, typical pathology [24].

Moreover, magnification is manually increased to 100x, and the step is repeated for 200x and 400x in order [16]. Images in BreaKHis are acquired in 24-bit RGB format with a dimension of 700x460 pixels [16]. Figure 1 shows samples from the BreaKHis dataset at the four magnification factors [16]. Interested, one can find a more detailed description of the BreaKHis dataset in Spanhol et al. (2016) ([16], [60]).

The BreaKHis dataset contains 7909 breast cancer images within two classes with subclasses in its first version [16]. The main classes are benign and malignant [16].

The benign image class contains 2440 images, and the malignant image class has 5429 images [60]. Each main category has four subclasses, and these subclasses are benign and can be named; Adenosis (A), Fibroadenoma (F), Tubular Adenoma (TA), and Phyllodes Tumor (PT) [34]. The other four subclasses, which are the subclasses of malignant, are Ductal Carcinoma (DC), Lobular Carcinoma (LC), and Mucinous Carcinoma (MC), and Papillary Carcinoma (PC) [34].

This paper only focuses on the subclasses of the malignant class; all benign subclasses are taken as benign. Moreover, malignant is separated into subclasses making five classes in total. For that reason, Figure 2 shows benign types with different magnifications, but Figure 3 shows malignant subclasses at a specific resolution. In addition to what was discussed above, Table 1 and Figure 4 show the benign image count for each resolution 40x, 100x, 200x, and 400x, and the four different subclasses of malignant.

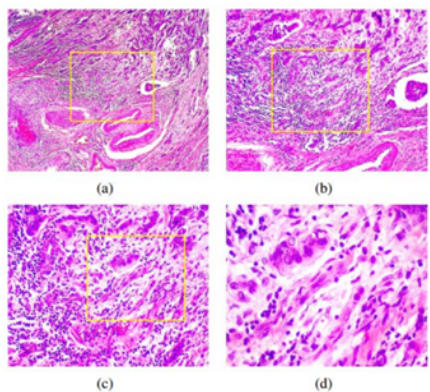

Fig. 1: Breast cancer tumors with different magnifications of the microscope: (a) 40x, (b) 100x, (c) 200x, and (d) 400x. The highlight rectangle represents the area of interest seen at higher magnification [18].

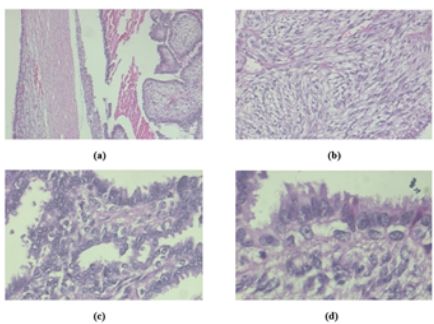

Fig. 2: Breast cancer tumors with different magnifications of the microscope, belonging to benign class: (a) 40x, (b) 100x, (c) 200x, and (d) 400x [18].

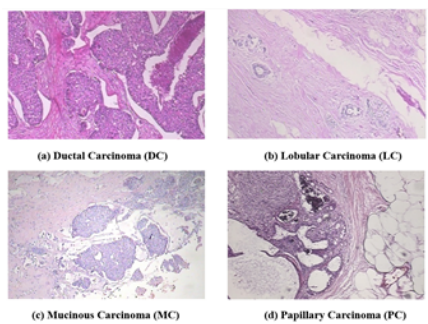

Fig.3: The four subclasses of a malignant class, at a 40X scale [18].

Table 1. Image distribution by magnification factor and histological subtypes that are used.

| Magnification | Benign | Malignant | | | | Total |
|---|---|---|---|---|---|---|
| | | Ductal Carcinoma (DC) | Lobular Carcinoma (LC) | Mucinous Carcinoma (MC) | Papillary Carcinoma (PC) | |
| 40x | 625 | 864 | 156 | 205 | 145 | **1995** |
| 100x | 644 | 903 | 170 | 222 | 142 | **2081** |
| 200x | 623 | 896 | 163 | 196 | 135 | **2013** |
| 400x | 588 | 788 | 137 | 169 | 138 | **1820** |
| **Total Images** | **2480** | **3451** | **626** | **792** | **560** | **7909** |

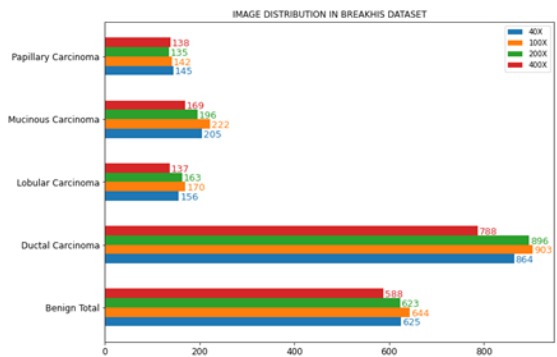

Fig. 4: Distribution of the total benign cancer images and four malignant subclasses for 40x, 100x, 200x, and 400x magnifications.

*B. Data Preprocessing*

For multiclass classification, 40x magnified images have been used. One of the main challenges in classifying the Histopathology images is having different coloring in photos. Even though in the BreakHis dataset, all images are H&E stained, there is a significant difference stemming from various reasons, like using different machines to obtain images, the difference between stain solutions depending on where it has been produced, and due to lab protocols [61].

Deep Learning models can learn general patterns from various colors, but it requires massive amounts of data, which is lacking in BreakHis. Therefore stain normalization is necessary to transform all the input into the same distribution. In 2001, Reinhard et al. [62] proposed a general statistical approach to style transfer for the target image based on the source image that can also be used for the histopathology images.

Macenko et al. [63] have used correction and variation of histopathology images based on their optical densities, separating the H and E stans. More advanced techniques have been developed recently than direct mathematical transformation using GAN models (Runz et al., 2021) [64].

Among these techniques, we choose the less expensive algorithm, which is also structure-preserving. A method is based on Vahadane et al. 2015 paper called "Structure-preserving stain normalization for histopathology images," where they learn the structure of the target image and normalize all the other images based on the target's structure; Figure 5 shows how the images result in the same color distribution after the normalization [61].

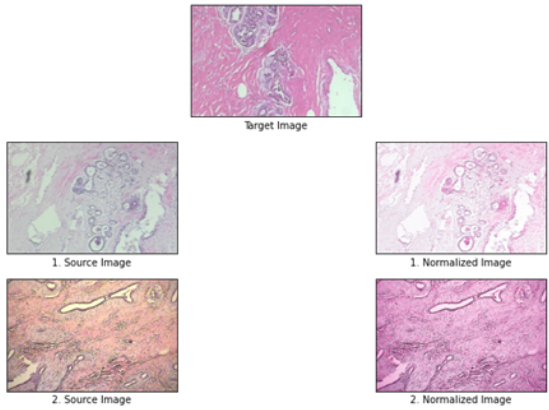

Fig. 5: Top image is the target whose structure is applied, left are examples of source images, and right are corresponding stain normalized images.

The method takes the target image's color space as a base and converts source images into stain normalized ones by using sparse non-negative matrix factorization (SNMF). Therefore, target and source images are converted from RGB to optical densities, then color change is applied to the source image and converted back to RGB [60]. The formula for calculating relative optical density - V is given as follows:

$$V = log(\frac{I_0}{I}) \quad (1)$$

I is the matrix of RGB image intensities, and I_0 is the illuminating light on the sample image. The relation between I and I_0 is:

$$I = I_0 * e^{(-W*H)} \quad (2)$$

Where W is the stain color matrix, columns correspond to the RGB channel of each stain, and H is called the concentration matrix, whose rows represent the total amount of the stained tissue. Then by substituting the equation 2 into 1, the relative optical density formula becomes as follows:

$$V = W * H \quad (3)$$

After having all the images normalized, for 40x magnification, they were split into train and test data with the ratio of 80:20, respectively. Afterward, training data is divided into train and validation by 80:20 percent; hence 64 % of the whole data is used for training, 16 % for validation, and 20 % for testing.

*C. Class Level Data Augmentation*

According to *Table 1*, in 40x, the imbalance ratio between Ductal and Papillary Carcinoma is almost 6:1 and considering that there are only 145 images for papillary, it can be viewed as a severe imbalance issue and lack of data. The situation is similar for the Lobular and Mucinous Carcinoma too. Therefore, we focus on enriching each class according to being a majority or minority class rather than concentrating on modeling, improving the overall score, and solving the imbalance problem. It is also proven that applying the same data augmentation for the whole dataset increases model performance on average results, and fine-tuning pre-trained models can surpass the SOTA. However, the classes that are not label preserving are being negatively affected. Since all the malignant types are essential in breast cancer detection to save lives, the main goal should not be outperforming previous works on average results or accuracy but providing a reliable model for all classes, even though there can be some trade-offs for test results.

For Benign images, there are 400 images in training data; for each image, seven times data augmentation has been applied, as shown in the *Algorithm 1*: horizontal and vertical flip, center crop with size 224x224, 30 and 60 degrees anticlockwise rotation, and twice random shearing with a range between 0.3 and 0.5 resulting in 3200 images in total.

```
Algorithm 1: Image Augmentation for Benign Class
    input: an array of training images (in_images)
    output: an array of augmented training images (out_images)
    function DataAugmentor(in_images)
1       n ← number of images in the input array
2       i ← 0
3       while i < n do
4           img ← read image i from in_images
5           img-HF ← RandomHorizontalFlip(p=1)(img)
6           img-VF ← RandomVerticalFlip(p=1)(img)
7           img-CC ← CenterCrop(size=(224,224))(img)
8           img-rot-30 ← Rotate((angle=30)(img)
9           img-rot-60 ← Rotate((angle=60)(img)
10          img-AT1, img-AT2 ← RandomAffine(shear=(0.3, 0.5), degrees=(0))(img) – 2 times
11          add all the images to the out_images array
12          i ← i + 1
13      end
14  end
```

Algorithm 1: data augmentation used for benign class

There are 553 images in ductal carcinoma, and each image has been augmented five times, as given in *Algorithm 2*: horizontal and vertical flip, center crop by the size of 224x224, rotation 30 degrees, and random shearing with a range between 0.3 and 0.5. As a result, 2765 new images are produced.

```
Algorithm 2: Image Augmentation for Ductal Carcinoma Class
    input: an array of training images (in_images)
    output: an array of augmented training images (out_images)
    function DataAugmentor(in_images)
1       n ← number of images in the input array
2       i ← 0
3       while i < n do
4           img ← read image i from in_images
5           img-HF ← RandomHorizontalFlip((p=1))(img)
6           img-VF ← RandomVerticalFlip((p=1))(img)
7           img-CC ← CenterCrop(size=(224, 224))(img)
8           img-rot-30 ← Rotate(angle=30)(img)
9           img-AT ← RandomAffine(degrees=(0), shear=(0.3, 0.5))(img)
10          add all the images to the out_images array
11          i ← i + 1
12      end
13  end
```

Algorithm 2: data augmentation for ductal carcinoma

Lobular Carcinoma initially contains 100 images in training data; therefore, it has been heavily augmented. As shown in *Algorithm 3*: each image has been extended 30 times: horizontal and vertical flip, five crops from upper left, upper right, bottom left, bottom right, and center by the size of 224x224, five times color jittering with brightness - 0.5, hue - 0.3 and saturation 0.4, 5 times random affine with degrees between 30 and 70 degrees, translation between 0.1 and 0.4, and 6 times random shearing with range 0.1 and 0.4 have been used on original images. The same shearing has been applied once for each of the five crop images, horizontally and vertically flipped images; as a result, 3000 new images are produced.

```
Algorithm 3: Image Augmentation for Lobular Carcinoma Class
    input: an array of training images (in_images)
    output: an array of augmented training images (out_images)
    function DataAugmentor(in_images)
1       n ← number of images in the input array
2       i ← 0
3       while i < n do
4           img ← read image i from in_images
5           img-HF ← RandomHorizontalFlip((p=1))(img)
6           img-VF ← RandomVerticalFlip((p=1))(img)
7           img-FC1, img-FC2, img-FC3, img-FC5, img-FC5 ← FiveCrop(size=(224,224))(img)
8           img-CJ1, img-CJ2, img-CJ3, img-CJ4, img-CJ5 ← ColorJitter(brightness=0.5, hue=0.3, saturation=0.4)(img) – 5 times
9           img-RA1, img-RA2, img-RA3, img-RA4, img-RA5 ← RandomAffine(degrees=(30, 70), translate=(0.1, 0.4))(img) – 5 times
10          img-FC1-CJ ← ColorJitter(img-FC1)
11          img-FC2-CJ ← ColorJitter(img-FC2)
12          img-FC3-CJ ← ColorJitter(img-FC3)
13          img-FC4-CJ ← ColorJitter(img-FC4)
14          img-FC5-CJ ← ColorJitter(img-FC5)
15          img-RS1, img-RS2, img-RS3, img-RS4, img-RS5, img-RS6 ← RandomAffine(degrees=0, shear=(0.1, 0.4))(img) – 6 times
16          img-HF-RS ← RandomAffine(degrees=0, shear=(0.1, 0.4))(img-HF)
17          img-VF-RS ← RandomAffine(degrees=0, shear=(0.1, 0.4))(img-VF)
18          add all the images to the out_images array
19          i ← i + 1
20      end
21  end
```

Algorithm 3: data augmentation for lobular carcinoma

In Mucinous Carcinoma for 40x magnification, there are 132 images in the training set. Each image has been augmented 23 times as given in *Algorithm 4*: random and horizontal flip, five crops from upper left, upper right, bottom left, bottom right, and center by the size of 224x224, three times color jittering with brightness - 0.5, hue - 0.3 and saturation 0.4, 5 times random affine with degrees between 30 and 70 degrees, translation between 0.1 and 0.4. The same shearing has been used horizontally and vertically flipped images on the upper left, center, and bottom right crops. As a result, 3036 new images have been generated.

```
Algorithm 4: Image Augmentation for Mucinous Carcinoma Class
    input: an array of training images (in_images)
    output: an array of augmented training images (out_images)
    function DataAugmentor(in_images)
1       n ← number of images in the input array
2       i ← 0
3       while i < n do
4           img ← read image i from in_images
5           img-HF ← RandomHorizontalFlip(p=1)(img)
6           img-VF ← RandomVerticalFlip(p=1)(img)
7           img-FC1, img-FC2, img-FC3, img-FC5, img-FC5 ← FiveCrop(size=(224, 224))(img) – 5 times
8           img-CJ1, img-CJ2, img-CJ3 ← ColorJitter(brightness=0.5, hue=0.3, saturation=0.4)(img) – 3 times
9           img-RA1, img-RA2, img-RA3, img-RA4, img-RA5 ← RandomAffine(degrees=(30, 70), translate=(0.1, 0.4))(img) – 5 times
10          img-FC1-CJ ← ColorJitter(img-FC1)
11          img-FC3-CJ ← ColorJitter(img-FC3)
12          img-FC5-CJ ← ColorJitter(img-FC5)
13          img-RS1, img-RS2, img-RS3 ← RandomAffine(degrees=0, shear=(0.1, 0.4))(img) – 3 times
14          img-HF-RS ← RandomAffine(degrees=0, shear=(0.1, 0.4))(img-HF)
15          img-VF-RS ← RandomAffine(degrees=0, shear=(0.1, 0.4))(img-VF)
16          add all the images to the out_images array
17          i ← i + 1
18      end
19  end
```

Algorithm 4: data augmentation for mucinous carcinoma

Papillary Carcinoma contains 93 images in the training set; therefore, heavy data augmentation has been used 33 times for each image as given in *Algorithm 5*: horizontal and vertical flip, five crops from upper left, upper right, bottom left, bottom right, and center by the size of 224x224, five times color jittering with brightness - 0.5, hue - 0.3 and saturation 0.4, 5 times random affine with degrees between 30 and 70 degrees, translation between 0.1 and 0.4, and 6 times random shearing with range 0.1 and 0.4 have been used on original images.

Same color jittering has been used once for each of the five cropped images, and the same shearing has been applied once for upper left, upper right, and center crops, horizontally and vertically flipped images. Three thousand sixty-nine new images have been produced.

```
Algorithm 5: Image Augmentation for Papillary Carcinoma Class
    input: an array of training images (in_images)
    output: an array of augmented training images (out_images)
    function DataAugmentor(in_images)
1       n ← number of images in the input array
2       i ← 0
3       while i < n do
4           img ← read image i from in_images
5           img-HF ← RandomHorizontalFlip(p=1)(img)
6           img-VF ← RandomVerticalFlip(p=1) (img)
7           img-FC1, img-FC2, img-FC3, img-FC5, img-FC5 ← FiveCrop(size=(224, 224))(img) – 5 times
8           img-CJ1, img-CJ2, img-CJ3, img-CJ4, img-CJ5 ← ColorJitter(brightness=0.5, hue=0.3, saturation=0.4)(img) – 5 times
9           img-RA1, img-RA2, img-RA3, img-RA4, img-RA5 ← RandomAffine(degrees=(30, 70), translate=(0.1, 0.4))(img) – 5 times
10          img-FC1-CJ ← ColorJitter(img-FC1)
11          img-FC2-CJ ← ColorJitter(img-FC2)
12          img-FC3-CJ ← ColorJitter(img-FC3)
13          img-FC4-CJ ← ColorJitter(img-FC4)
14          img-FC5-CJ ← ColorJitter(img-FC5)
15          img-RS1, img-RS2, img-RS3, img-RS4, img-RS5, img-RS6 ← RandomAffine(degrees=0, shear=(0.1, 0.4))(img) – 6 times
16          img-HF-RS ← RandomAffine(degrees=0, shear=(0.1, 0.4))(img-HF)
17          img-VF-RS ← RandomAffine(degrees=0, shear=(0.1, 0.4))(img-VF)
18          img-FC1-RS ← RandomAffine(degrees=0, shear=(0.1, 0.4))(img-FC1)
19          img-FC2-RS ← RandomAffine(degrees=0, shear=(0.1, 0.4))(img-FC2)
20          img-FC3-RS ← RandomAffine(degrees=0, shear=(0.1, 0.4))(img-FC3)
21          add all the images to the out_images array
22          i ← i + 1
23      end
24  end
```

Algorithm 5: data augmentation for papillary carcinoma

For the pre-processing step, the algorithms mentioned above are used for different classes' training images described to have a balanced dataset. After the data augmentation, making each class have approximately 3200 images solved the data imbalance problem by making each class have around 3200 images. The exact number of samples in the augmented training set can be seen in *Table 2*.

Table 2: Image distribution of training data after data augmentation

|  | Benign | Malignant | | | |
|---|---|---|---|---|---|
|  |  | Ductal Carcinoma (DC) | Lobular Carcinoma (LC) | Mucinous Carcinoma (MC) | Papillary Carcinoma (PC) |
| Number of images in training set | 400 | 553 | 100 | 132 | 93 |
| Data augmentation factor | x7 | x5 | x30 | x23 | x33 |
| Total images after augmentation | 3200 | 3318 | 3300 | 3168 | 3162 |

Examples of augmented images for the benign class have been given in *Figure 6*. After all the augmentation process was done, images were resized by 224x224, and the format changed from 224x224x3 to 3x224x224. As code has been implemented in Pytorch, images must first have channel dimensions based on the framework's requirements. Besides, all the images are normalized using equation *4* due to the conditions of the software tool before passing to the model, which is elaborately explained in part D.

$$x_n = \frac{x - \mu}{\sigma} \quad (4)$$

$x_n$ - normalized image channel, μ -mean value of pixels,

σ - standard deviation of the pixels for the channel

Note that mean values are 0.485, 0.456, 0.406 and standard deviations are 0.229, 0.224, 0.225 for R, G, and B channels, respectively.

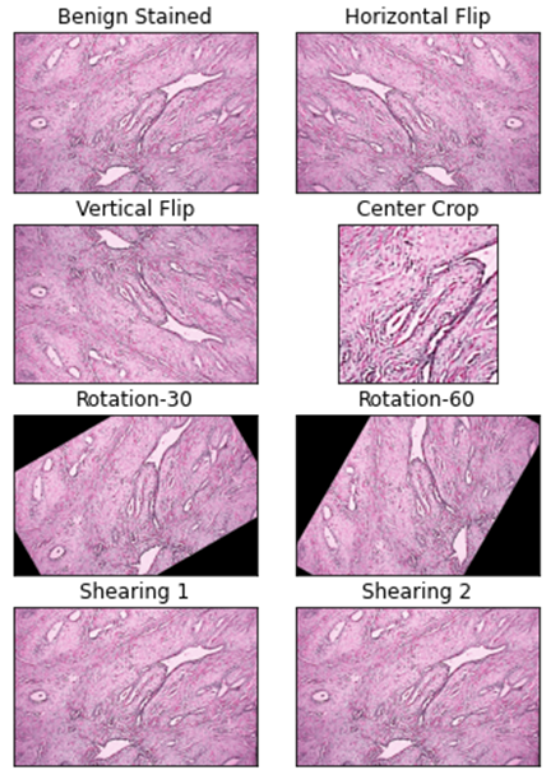

Fig. 6: example augmented data for benign class

### D. ViTNet and CNN Models

**Main architecture** - ViTNet used in this paper is based on the original Vision Transformers [65]. The structure of the ViTNet is given in *Figure 7*, similar to [65], by differing with the number of layers used as classifiers. A High-level overview of the architecture is as follows: use necessary image processing to prepare input data, pass the resultant inputs to the encoder block, and apply multilayer perceptron on produced feature maps to classify images. Encoder blocks play a role as feature extractors, the same as the encoder part of the "Attention is all you need" paper [66]. However, how the input is passed to the encoder differs, as inputs represent images, not words or sentences. Therefore, preparing input suitable for the transformer encoder is necessary. For this reason, all the images are divided into 16x16 patches, as images have been resized by 224x224, making 196 patches for a single image. Then, the sequence of the patches has been projected into a new space where they are concatenated with positional embeddings that help to get the advantage of the sequence relation, in other words, local similarities of the neighborhood pixels.

The Transformer Encoder consists of 12 Encoder blocks, as shown in *Figure 8*. First, embedded patches are normalized across each layer, and then the result is passed to the Multihead Attention layer.

As Multihead attention uses the self-attention mechanism [66], an additional learnable weight matrix is used to produce keys, queries, and values.

Let $I \in R^{NxD}$ be the result of the layer normalization and $W_{QKV} \in R^{Dx3D_k}$, our first weight matrix in the Multihead Attention layer. Then, by multiplying input $I$ by the weight matrix, we get $I_0 \in R^{Nx3D_k}$, and $I_0$ is separated into three matrices along the dimension $3D_k$ producing query, key, and value - Q, K, V $\in R^{NxD_k}$, as shown in *Equation 5*. Using Scaled Dot Product Attention - SDPA results in a matrix with the same dimensionality as the value matrix (*Equation 6*). This process is repeated by h=12 times with different learnable parameters, where h represents the number of heads. After the Scaled Dot Product is done, the resultant h=12 matrices are concatenated, $cat[SDPA_1, ..., SDPA_h] \in R^{Nxh*D_k}$, and produced matrix is multiplied by the second weight matrix - $W_O \in R^{h*D_k xD}$, that gives us the final matrix for the Multihead Attention layer - $MHA \in R^{NxD}$ that has the exact dimensionality with the $I \in R^{NxD}$ (*Equation 7*).

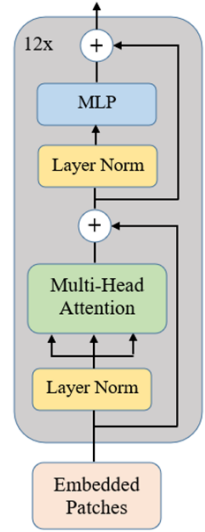

Fig. 8: Transformer Encoder Block

$$Q, K, V = I * W_{QKV}, \quad W_{QKV} \in R^{Dx3D_k}; \; Q, K, V \in R^{NxD_k} \quad (5)$$

$$SDPA = Softmax(\frac{Q*K^T}{\sqrt{D_k}}) * V, \quad SDPA \in R^{NxD_k} \quad (6)$$

$$MHA = cat[SDPA_1, ..., SDPA_h] * W_O,$$
$$W_O \in R^{h*D_k xD}; \; MHA \in R^{NxD} \quad (7)$$

By using the residual connection [13], embedded patches are added to the output of the Multihead Attention layer, producing the first residual result - R1. Then R1 is passed to the layer normalization, and output is given to the MLP layer (*Equation 8*) that represents the position-wise feed-forward network in [66], but with a slight difference of using GELU nonlinearity [67] (*Equation 9*) between fully connected layers, instead of ReLU.

$$MLP(x) = GELU(x * W_1 + b_1) * W_2 + b_2 \quad (8)$$

$$GELU(x) = 0.5 * x * (1 + erf(\frac{x}{\sqrt{2}})) \quad (9)$$

$$erf(\frac{x}{\sqrt{2}}) \cong tanh(\sqrt{\frac{2}{\pi}} * (x + 0.044715 * x^3)) \quad (10)$$

Note that *Equation 10* is the approximation used for the calculation of the *erf* function and dimensionality of the weight matrices in *Equation 8* is as followers:

$$W_1 \in R^{768x3072}; \; W_2 \in R^{3072x768}$$ $b_1$ and $b_2$ represent biases for each linear layer.

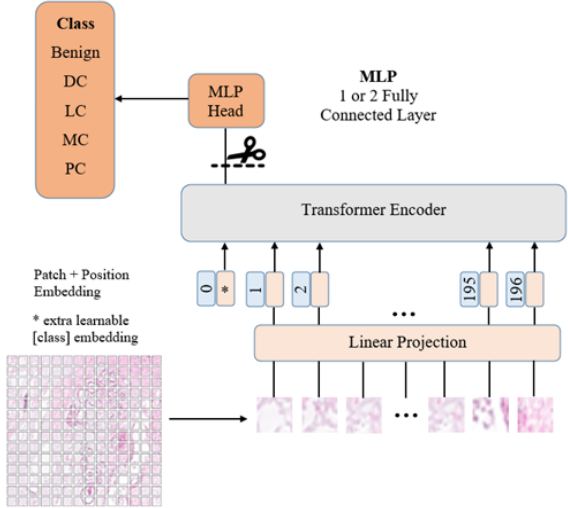

Fig. 7: VitNet Architecture

Finally, the first residual result - $R_1$ is added to the MLP layer's output, which finishes the first encoder block. Twelve encoder blocks have been used in the architecture, meaning each encoder's outcome is passed to the next encoder block, and the final block produces a matrix of output dimensionality

768.

For classification of the breast cancer images, Vision Transformer with 12 encoder blocks using 12 heads in Multihead Attention, 3072 internal layer size, 768 input, and output layer size in MLP layer and patch size of 16x16, pre-trained on Imagenet dataset has been used. As given in *Figure 7*, the classifier part has been removed after the training on the Imagenet, and learned weights in the encoder blocks and learned embedding have been kept but frozen to avoid further training while transfer learning on the BreakHis dataset. Then, 1 or 2 linear layers have been added at to top to construct ViTNet for the multiclass classification. For ViTNet with one learnable layer at the top, a fully connected layer with input size 768, output size 5, followed by a softmax layer, has been used. The structure of the ViTNet with two learnable layers is given in *Figure 9*.

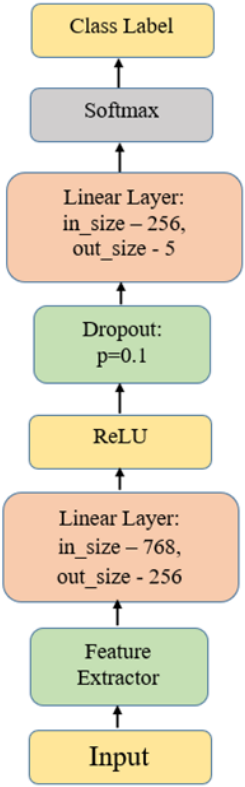

Fig. 9: ViTNet Architecture

VitNet architecture uses transformer encoders for feature extraction and 2 linear layers for classification. Between linear layers, ReLU activation and dropout have been used to learn nonlinear representation and avoid overfitting, respectively. At the last layer, softmax is used to convert the results into probability distribution across the 5 classes, where the class with the highest probability is considered the predicted label.

To compare our results with the current state-of-the-art methods in multiclass breast cancer classification which will be discussed in detail in Section IV, we also use Alexnet, VGG16, and Inception-v3 that have been pre-trained on Imagenet. By removing the head classifier, we use the body part as a feature extractor by freezing the layers and adding one linear layer on the top with 5 units for classification. In the next section, we discuss the evaluation metrics we used, and how our proposed ViTNet performs compared to the fully CNN-based models.

## 4 RESULTS

*Training*

The models created, ViTNet with 1 Fully Connected Layer and ViTNet with 2 Fully Connected Layers, are evaluated in the testing dataset for benign, ductal carcinoma, lobular carcinoma, mucinous carcinoma, and papillary carcinoma classes separately, as well as on average and compared to the Alexnet, VGG16, and Inception-v3 models. All the training has been done using Python3 language, and PyTorch library on Google Colab Pro with Tesla T4 GPU. Training is operated for 20 epochs with batch size 64, Adam optimizer [72] with a learning rate of 0.001, and cross-entropy as a loss function.

*Performance Metrics*

Unlike similar studies, patient-based performance is not evaluated in this work since the possible higher results can be filtered out to achieve generalization.

The accuracy metric alone is not sufficient to express the generalized performance of a model for an imbalanced dataset, as is investigated in the study of Bekkar et al. in 2013 [68] since the model might label most of the data as the majority class and still be able to have a high accuracy value. Therefore, for the evaluation of the network, the following metrics are chosen: Precision, the ratio of true labeled positives to all data marked by the model as positive; Recall, the percentage of true labeled positives to all real positives in the dataset and F1-score, the harmonic mean of precision and recall; Specificity, the ratio of the true-negative labeled classes to all real negatives in the dataset; False Positive Rate, how often our model gives false alarms, that is opposite of the Specificity; False Negative Rate, how often our model misses the actual patients, which is the opposite of the recall. The formulas can be seen in Equations 11, 12, 13, 14, 15, and 16, respectively.

$$Precision = \frac{TP}{TP + FP} \quad (11)$$

$$Recall = \frac{TP}{TP + FN} \quad (12)$$

$$F1 - score = 2 \times \frac{Precision \times Recall}{Precision + Recall} \quad (13)$$

$$Specificity = \frac{TN}{TN + FP} \quad (14)$$

$$FPR = \frac{FP}{FP + TN} \quad (15)$$

$$FNR = \frac{FN}{FN + TP} \quad (16)$$

TP - True Positive, FP - False Positive, TN - True Negative, FN - False Positive, FPR - False Positive Rate or Fall Out, FNR - False Negative Rate or Miss Rate

Evaluation is done using the Scikit-Learn library for Python.

In the 2014 work of Zachariah et al.[69], the lift metric is recommended as it examines how much better a classifier performs relative to random chance. The lift metric is given in Equation 17.

$$Lift = (\frac{TP}{TP+FP})/(\frac{TP+FN}{TP+FP+TN+FN}) \quad (17)$$

To compare the performance of the created model with the existing state-of-the-art neural networks, Alexnet[70], InceptionV3[71], and VGG-16[71] are evaluated in their fine-tuned versions for evaluation on the BreaKHis dataset. Current best results have been achieved by using pre-trained weights of these models on the ImageNet dataset, which are used as feature extractors. Linear layers have replaced top layers with learnable parameters to classify images into five classes. The results for each class can be seen in Tables 3, 4, 5, 6, and 7, respectively, which show that the proposed models in this work match the state-of-art performances for this five-class classification task.

Table 3: Evaluation of different models for benign class on 40X magnification testing set with respect to different performance metrics

|  | Precision | Recall | F1-score | Specificity | FPR | FNR | Lift Score |
|---|---|---|---|---|---|---|---|
| Alexnet | 0.56 | 0.42 | 0.48 | 0.85 | 0.15 | 0.58 | 1.77 |
| Inception-v3 | 0.67 | 0.38 | 0.49 | 0.91 | 0.09 | 0.62 | 2.12 |
| VGG-16 | 0.65 | 0.37 | 0.47 | 0.91 | 0.09 | 0.63 | 2.05 |
| ViT with 1 Fully Connected Layer | 0.75 | **0.64** | **0.69** | 0.90 | 0.10 | **0.36** | 2.37 |
| ViT with 2 Fully Connected Layer | 0.81 | 0.56 | 0.67 | **0.94** | **0.06** | 0.44 | **2.57** |

An essential measure for tissue classification is the false-positive rate, especially in the benign class, since it represents the errors that occurred by labeling a malignant tissue as benign. A misdiagnosed cancerous tissue may be a fatal error since it will delay the specific cancer treatment. As it can be seen in Table 3, the proposed models in this work decrease the false-positive rate to 3%. Further should improve the performance of this metric.

Table 4: Evaluation of different models for ductal carcinoma class on 40X magnification testing set concerning different performance metrics

|  | Precision | Recall | F1-score | Specificity | FPR | FNR | Lift Score |
|---|---|---|---|---|---|---|---|
| Alexnet | 0.63 | 0.56 | 0.59 | 0.74 | 0.26 | 0.44 | 1.43 |
| Inception-v3 | 0.52 | 0.41 | 0.46 | 0.70 | 0.30 | 0.59 | 1.19 |
| VGG-16 | 0.55 | 0.59 | 0.57 | 0.63 | 0.37 | 0.41 | 1.26 |
| ViT with 1 Fully Connected Layer | **0.72** | 0.82 | **0.77** | **0.75** | 0.25 | 0.18 | **1.64** |
| ViT with 2 Fully Connected Layer | 0.65 | **0.86** | 0.75 | 0.64 | 0.36 | **0.14** | 1.49 |

As seen in Table 4, the F1 score is increased up to 18% in the proposed model. Since ductal carcinoma is the most common sub-type, the false positive rate is an important metric. If the data gets labeled as this majority class, minority classes may be left undiagnosed. The proposed models match with the best false positive rate among other models.

Table 5: Evaluation of different models for lobular carcinoma class on 40X magnification testing set with respect to different performance metrics

|  | Precision | Recall | F1-score | Specificity | FPR | FNR | Lift Score |
|---|---|---|---|---|---|---|---|
| Alexnet | 0.26 | 0.43 | 0.32 | 0.9 | 0.1 | 0.57 | 3.31 |
| Inception-v3 | 0.21 | 0.46 | 0.29 | 0.85 | 0.15 | 0.54 | 2.72 |
| VGG-16 | 0.25 | 0.43 | 0.32 | 0.89 | 0.11 | 0.57 | 3.2 |
| ViT with 1 Fully Connected Layer | 0.67 | **0.54** | **0.59** | **0.98** | **0.02** | **0.46** | 8.45 |
| ViT with 2 Fully Connected Layer | **0.70** | 0.46 | 0.56 | **0.98** | **0.02** | 0.54 | **8.91** |

Since lobular carcinoma tumor has a critical placement in lymph, identifying all positive samples is essential. As seen in Table 5, the proposed models can increase recall value up to 8%. The proposed models also have high precision values compared to others. There is a 27% significant improvement in the F1 score.

Table 6: Evaluation of different models for mucinous carcinoma class on 40X magnification testing set with respect to different performance metrics

|  | Precision | Recall | F1-score | Specificity | FPR | FNR | Lift Score |
|---|---|---|---|---|---|---|---|
| Alexnet | 0.33 | 0.29 | 0.31 | 0.93 | 0.07 | 0.71 | 3.19 |
| Inception-v3 | 0.24 | 0.23 | 0.24 | 0.92 | 0.08 | 0.77 | 2.33 |
| VGG-16 | 0.22 | 0.16 | 0.19 | 0.93 | 0.07 | 0.84 | 2.16 |
| ViT with 1 Fully Connected Layer | **0.45** | **0.40** | **0.42** | 0.94 | 0.06 | **0.60** | 4.34 |
| ViT with 2 Fully Connected Layer | **0.45** | 0.36 | 0.40 | **0.95** | **0.05** | 0.64 | **4.36** |

Mucinous carcinoma is a minority class in the dataset. However, the proposed models may increase the F1 score up to 11% percent as can be seen in Table 6.

Table 7: Evaluation of different models for papillary carcinoma class on 40X magnification testing set with respect to different performance metrics

|  | Precision | Recall | F1-score | Specificity | FPR | FNR | Lift Score |
|---|---|---|---|---|---|---|---|
| Alexnet | 0.12 | 0.29 | 0.17 | 0.86 | 0.14 | 0.71 | 1.9 |
| Inception-v3 | 0.15 | 0.49 | 0.24 | 0.82 | 0.18 | 0.51 | 2.44 |
| VGG-16 | 0.17 | 0.38 | 0.23 | 0.88 | 0.12 | 0.62 | 2.68 |
| ViT with 1 Fully Connected Layer | 0.34 | 0.40 | 0.37 | 0.95 | 0.05 | 0.6 | 5.36 |
| ViT with 2 Fully Connected Layer | **0.40** | **0.42** | **0.41** | **0.96** | **0.04** | **0.58** | **6.25** |

The papillary carcinoma class is one of the minority classes in the dataset. Therefore, the F1-score metric, harmonic mean of recall, and precision values are at most important. As it can be seen in Table 7, the proposed models have increased the performance up to 17%.

Table 8: Evaluation of different models on 40X magnification testing set with respect to different performance metrics

|  | Accuracy | Macro Averaged Precision | Weighted Averaged Precision | Macro Averaged Recall | Weighted Averaged Recall | Macro Averaged F1-score | Weighted Averaged F1-score |
|---|---|---|---|---|---|---|---|
| Alexnet | 0.46 | 0.38 | 0.51 | 0.40 | 0.46 | 0.37 | 0.48 |
| Inception-v3 | 0.28 | 0.33 | 0.32 | 0.26 | 0.28 | 0.26 | 0.27 |
| VGG-16 | 0.47 | 0.41 | 0.53 | 0.43 | 0.47 | 0.39 | 0.47 |
| ViT with 1 Fully Connected Layer | **0.67** | 0.58 | **0.67** | **0.56** | **0.67** | **0.57** | **0.67** |
| ViT with 2 Fully Connected Layer | 0.66 | **0.60** | **0.67** | 0.53 | 0.66 | 0.56 | 0.65 |

After evaluating each model's performance on different classes, weighted and macro averaged precision, recall, and F1 scores are calculated along with the accuracy metric to measure the overall performance. The results can be seen in Table 8, which again matches the state-of-art performance with an increase of a minimum of 10% and a maximum of 20%.

Table 9: Number of total and training parameters of the evaluated models

|  | Number of Total Parameters | Number of Trainable Parameters | Training Time |
|---|---|---|---|
| Alexnet | 57B | 20485 | 0.87h |
| Inception-v3 | 21.8B | 10245 | 1.81h |
| VGG-16 | 134.3B | 20485 | 1.46h |
| ViT with 1 Fully Connected Layer | 85.8B | 3845 | 1.55h |
| ViT with 2 Fully Connected Layer | 85.9B | 198149 | 1.57h |

Table 9 summarizes the evaluated models in terms of total and trainable parameters. As the resultant tables show, ViT with 1 Fully Connected Layer gives one of the best performances even though it only has 3845 trainable parameters. Training times of the models are also given in the Table 9, as it was mentioned in Training section.

## 5. CONCLUSION

This study proposes ViT-Net, a multi-classifier model for detecting breast cancer from histopathological images. Differentiating from previous deep learning-based methods, the model focuses on classifying the tissues not only as benign or malignant, the further sub-classification, if the sample is malignant since the sub-class label of the malignant case, may reveal significant information about the position of cancer, the stage and the progression of cancer, the possibility of metastases, and specific treatment procedures that the doctors should follow.

The evaluation is performed on the BreaKHis dataset, and four sub-classes under the benign class are kept together. In contrast, the malignant class is divided into four labels, which results in five classes in total. Different data augmentation methods pre-process the imbalanced dataset. Vision transformers with versions of one and two fully connected layers, which are not widely investigated methods for multi-classification in histopathological images, are proposed as the model.

As the proposed ViTNet models are evaluated and compared with the best performances achieved in breast cancer classification, it is seen that they improve other models' multi-classification performance by 10-20%. Also, the classification of cancerous tissues may be crucial depending on the sub-type, and it is seen that the ViTNet models increase the class-based F1 scores each by 11-23%. As is seen in the tables and graphs discussed in the results section, ViTNet versions 1 and 2 are by far the best models in terms of all performance metrics in multi-classification.

In future work, other histopathological breast cancer diagnosis datasets can be used to investigate the proposed models' performance. In this work, encoder blocks are frozen to avoid further learning and used as feature extractors in this study. For future improvement, it is planned to allow the last layers to learn on the BreakHis dataset and fine-tune the number of layers to be a freezer. Self-Supervised Vision Transformer is planned to be used to compare the results, such as DINO [72] and MobileVit [73], to get a more lightweight model that can be transferred into less powerful devices. Additionally, augmentation techniques for each class can be tuned in a hierarchical manner: first step involves selection features for each class, next step is about tuning of corresponding hyperparameter for each augmentation technique. This method is more costly, but more promising in terms of results.

ACKNOWLEDGMENT

We thank to Muhammed Burak BEREKETOGLU on scientific assistance on medical information.